\documentclass[twocolumn,showpacs,preprintnumbers,amsmath,amssymb]{revtex4}
\usepackage{tabularx,graphicx}\begin{document}
\newcommand{\beq}{\begin{equation}}
\newcommand{\eeq}{\end{equation}}
\newcommand{\beqn}{\begin{eqnarray}}
\newcommand{\eeqn}{\end{eqnarray}}
\newcommand{\bmath}{\begin{subequations}}
\newcommand{\emath}{\end{subequations}}
\newcommand{\rb}{\bar{r}}
\newcommand{\bk}{\bold{k}}
\newcommand{\bkp}{\bold{k'}}
\newcommand{\bq}{\bold{q}}
\newcommand{\bkb}{\bold{\bar{k}}}
\newcommand{\br}{\bold{r}}
\newcommand{\brp}{\bold{r'}}
\newcommand{\vp}{\varphi}

\title{Effect of orbital relaxation on the band structure of cuprate superconductors and implications for the superconductivity mechanism}
\author{J. E. Hirsch }
\address{Department of Physics, University of California, San Diego\\
La Jolla, CA 92093-0319}

\begin{abstract} 
Where the doped holes reside in cuprate superconductors has crucial implications for the understanding of the mechanism responsible for their high temperature
superconductivity. It has been generally assumed that doped holes reside in hybridized Cu $d_{x^2-y^2}$ - O $p\sigma$ orbitals in the $CuO_2$ planes,
based on results of density functional band structure calculations. Instead, we propose that doped holes in the cuprates reside in O $p\pi$ orbitals in the plane,
perpendicular to the $Cu-O$ bond, that are raised to the Fermi energy through local orbital relaxation, that is not taken into account in band structure
calculations that place the bands associated with these orbitals well below the Fermi energy. We use a dynamic Hubbard model to incorporate the orbital relaxation degree of freedom and find in exact diagonalization of a small $Cu_4O_4$  cluster
that holes will go to the O $p\pi$ orbitals for relaxation energies comparable to what is expected from atomic properties of oxygen anions. The
bandwidth of this band becomes significantly smaller than predicted by band structure calculations due to the orbital relaxation effect.
Within the theory of hole superconductivity the heavy hole carriers in this almost full band will pair and drive the system superconducting through lowering of their quantum kinetic energy.

 \end{abstract}
\maketitle 
\section{introduction}
The question of which atomic orbitals in the Cu-oxide superconductors host the carriers that drive the system superconducting was recognized as essential already in the early days of the
high $T_c$ cuprate era, since it is likely to play a key role in the understanding of the mechanism of superconductivity. Band structure calculations based on density functional theory (DFT)
yield a metallic ground state with a broad band of states with dominant Cu $d_{x^2-y^2}-O p\sigma$ orbital character crossing the
Fermi energy \cite{bs1,bs2,bs2p,bs3,bs4,bs5,bs6,bs7, bs8}. $O p \sigma$ orbitals are directed along the $Cu-O$ bond (see Fig. 1). Despite the fact that these calculations yield a metallic
rather than an  insulating state   in the undoped case, it has been generally accepted that holes doped into the insulating state go into this band and are the carriers
responsible for superconductivity.

 \begin{figure}
\resizebox{7.5cm}{!}{\includegraphics[width=7cm]{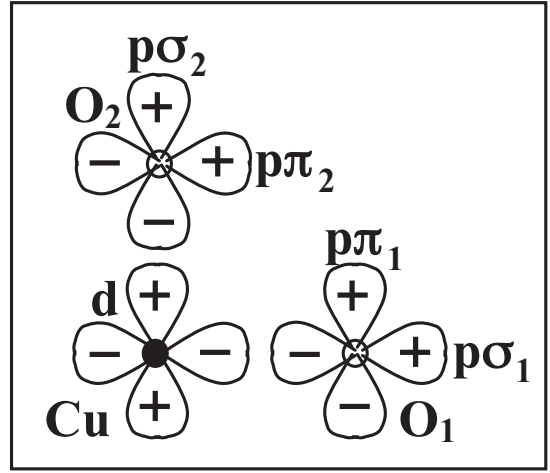}}
\caption {  Unit cell in the $Cu-O$ plane with $5$ electronic orbitals: $Cu$ $d_{x^2-y^2}$ and $O$ $p_\sigma$, $p_\pi$ orbitals. There are two oxygen atoms in the unit cell
denoted by $O_1$, $O_2$.}
\label{figure1}
\end{figure} 

The possibility that instead doped holes occupy planar oxygen orbitals pointing in direction perpendicular to the $Cu-O$ bond ($p \pi$ orbitals) was proposed early on by
Goddard and coworkers \cite{goddard} based on quantum chemical calculations. Also simple electrostatic considerations suggest that doped holes (positive carriers) should prefer
to occupy $O p \pi$ orbitals since they give rise to larger carrier concentration near the center of $Cu-O$ plaquettes (furthest away from the $Cu^{++}$ ions) where the
electrostatic potential is most negative. This was also found in detailed calculations of Madelung potentials \cite{adrian}.
However, it is generally believed that the strong hybridization between $O p \sigma$ and $Cu d_{x^2-y^2}$ orbitals along the $Cu-O$ bonds renders those orbitals more favorable
for hole doping as predicted by the DFT calculations.

Experimentally it is possible to ascertain that doped holes occupy planar $O$ orbitals ($p_{x,y}$ rather than $p_z$) from X-ray absorption \cite{x} and
electron energy loss spectroscopy \cite{eels} studies, but it is not possible to differentiate between $p \sigma$ and $p \pi$ holes.

The bands arising from direct overlap of oxygen orbitals are predicted by DFT theory to have a width of approximately $6$ to $7$ eV \cite{mms} and be located with the top at a distance of 
$1$ \cite{bs5} to $2.5$ $eV$ \cite{mms} below the 
Fermi energy, thus remaining completely filled by electrons and hence inert when the system is doped with holes.
In this paper we question this point of view and argue that it is in error, resulting from ignoring the important effect of {\it local orbital relaxation} of the filled oxygen orbitals when
an electron is removed. We argue that when this effect is taken into account the $O p \pi$ bands rise to the Fermi energy, and as a consequence
host the hole carriers responsible for superconductivity in the
cuprates \cite{hole1,hole2}.

\section{band structure}

As pointed out by Mattheiss and others\cite{bs1,bs2,bs2p,bs3,bs4,bs5,bs6,bs7,bs8}, the main features of the band structure of the cuprates obtained from density functional calculations can be reproduced with 
simple tight binding models  that include the $Cu$ $d_{x^2-y^2}$ and $O$ $p_x$, $p_y$ orbitals. Since superconductivity is clearly driven by transport in the planes,
we consider only   the $Cu$ $d_{x^2-y^2}$ and $O$ $p_x$, $p_y$ orbitals in the planes. There is one $Cu$ and two $O$ atoms in the unit cell which we denote by
$O_1$ and $O_2$. Figure 1 shows the orbitals schematically. The oxygen orbitals in the direction of the $Cu-O$ bond are denoted $p\sigma$ and the ones
perpendicular to the $Cu-O$ bond are denoted $p\pi$.

We denote the $d-p\sigma$ hopping amplitude by $t_d$, and the direct hopping amplitudes between oxygen orbitals by $t_1$ for $\pi-\pi$ or $\sigma-\sigma$ hopping
and $t_2$ for $\pi-\sigma$ hopping. Following estimates by McMahan et al \cite{mms} and Stechel and Jennison \cite{sj}
we take $t_1=0.65$, $t_2=0.35$ and $t_d=1.75$, all in $eV$. For site energies
we take $\epsilon_d=-5.2$, $\epsilon_{p\sigma}=-5.5$, $\epsilon_{p\pi}=-4.7$ $eV$. Because of electrostatics, $\epsilon_{p\pi}$ is higher than $\epsilon_{p\sigma}$.

 \begin{figure}
\resizebox{8.5cm}{!}{\includegraphics[width=7cm]{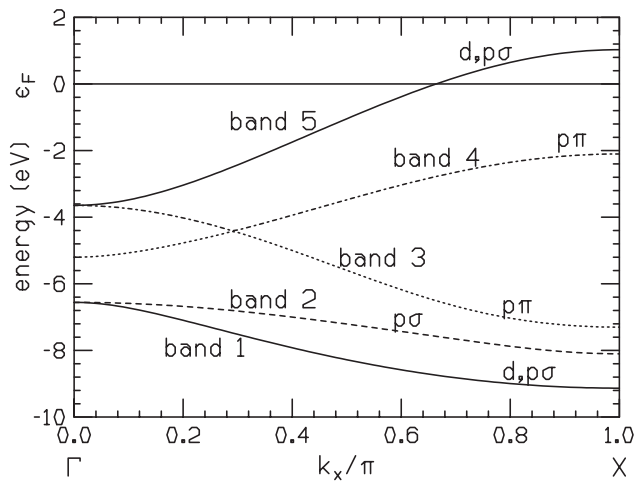}}
\caption {  Band structure in the $Cu-O$ planes in the $\Gamma-X$ direction 
($(0,0)$  to $(\pi,\pi)$) from a tight binding calculation with 5 orbitals per unit cell (see Fig. 1). Parameters used are
$t_{\pi\pi}=t_{\sigma\sigma}\equiv t_1=0.65$, $t_{\pi\sigma}\equiv t_2=0.35$, $t_{d\sigma}\equiv t_d=1.75$,   $\epsilon_d=-5.2$, $\epsilon_{p\sigma}=-5.5$, $\epsilon_{p\pi}=-4.7$   (see text).
}
\label{figure1}
\end{figure}

 \begin{figure}
\resizebox{9.0cm}{!}{\includegraphics[width=7cm]{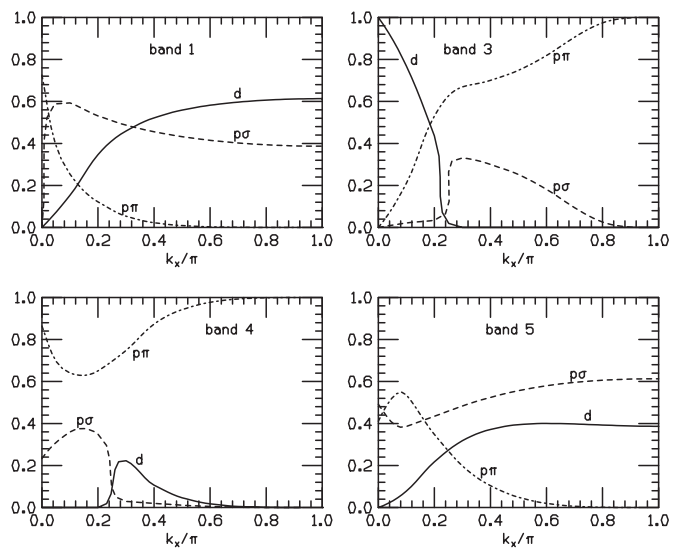}}
\caption {  Weights of the different orbitals in the band states. Bands 1 to 5 are numbered from lowest to highest energy.
Band 2 (not shown) has only contributions from $p\sigma$ (mostly) and $p\pi$, none from $d$, for all values of $k$.
}
\label{figure1}
\end{figure} 

The resulting band structure obtained by diagonalization of the 5x5 matrix of Bloch states is shown in Fig. 2 in the direction $(0,0)-(\pi,\pi)$, i.e. $\Gamma\rightarrow X$.
Figure 3 shows the weights of the different orbitals in the band states for bands 1, 3, 4 and 5 ordered from lowest
to highest energy. Band 2, that is not shown in Fig. 3, is mostly $p\sigma$, with some $p\pi$ contribution and  no contribution from the d orbitals.
The band of interest to us is band $4$, which is almost entirely of $p \pi$ character at its highest energies near the $X$ point  as shown in the 
lower left panel of Fig. 3.

The Fermi level corresponds to energy $0$. This tight binding 
band structure, extending from $-9.2eV$ to $1eV$,
resembles the main features of the band structures
obtained from density functional calculations \cite{bs1,bs2,bs2p,bs3,bs4,bs5,bs6,bs7,bs8}. The Fermi level cuts the $Cu-d_{x^2-y^2}-Op \sigma$ antibonding band that
extends from energy $-3.6eV$ to $1eV$, hence according
to this band structure
when the system is doped with holes they should occupy this band. This is the general consensus. The antibonding oxygen $p \pi$ band (band 4)
 is full and its top is approximately $2eV$ below the Fermi level, hence
it should remain full and   inert when the system is doped with holes according to this band structure.

There are two problems with this argument, the first one is well recognized but the second one is not.

The first problem is that the band structure in Fig. 2 does not reflect the fact that the undoped system is insulating. This is
attributed to the strong Coulomb repulsion of electrons in the $Cu-d_{x^2-y^2}$ orbital, which is argued to open up
a gap (Mott-Hubbard gap) when the band is half-filled, corresponding to the undoped case. Hence, band 5  in
Fig. 2 is argued to split into two, upper and lower Hubbard bands, when the system is undoped, 
with the Fermi level in the gap between the two bands rendering it insulating.

There have been various calculations performed using these ideas that take the Mott-Hubbard gap into account  \cite{bs3,bs7,mms,sj,svane,saw}. The general consensus is that
 doped holes still go into the
$Cu-d_{x^2-y^2}-Op \sigma$ band \cite{hyb,ann,saw2,rice} and are responsible for the transport in the underdoped through overdoped
regime. However, these calculations rely on approximations that are not necessarily well controlled.  

The second problem is that the antibonding $p\pi$ band (band 4 in Fig. 2), which comes to about $2eV$ below the Fermi level 
at the $X$ point, is assumed to be rigid. Here we argue that this assumption is incorrect and that in reality 
the energy of holes doped into this band is raised by several $eV$ by $orbital$ $relaxation$, and that as a result
doped holes will go into this band rather than into the $Cu-Op \sigma$ band.
In other words, the energetics of orbital relaxation, not reflected in the band structure shown in Fig. 2, makes
it easier to remove holes from  band 4 rather than from the band above it (band 5 in Fig. 2).

 \begin{figure}
\resizebox{9.0cm}{!}{\includegraphics[width=7cm]{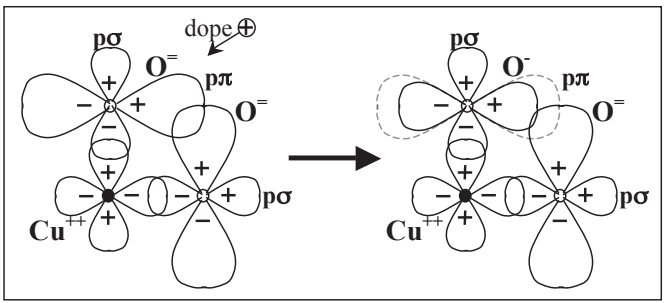}}
\caption { Atomic orbitals in the $CuO_2$ unit cell, for the undoped case (left) and when an electron is removed (hole is added) (right). If the hole goes
into the $p\pi$ orbital, it will shrink, as shown schematically with the dashed and full lines on the right panel of the figure.
}
\label{figure1}
\end{figure}

\section{orbital relaxation}

When an atomic orbital is doubly occupied, its size expands. This is certainly well known from atomic physics \cite{slater} but surprisingly its consequences are not
properly  taken into
account in band structure calculations nor in the many-body Hamiltonians commonly used for solids. 
In a series of papers we have argued that this effect is essential to understand the physics of electrons in electronic energy bands that are more than
half full \cite{dynh1,dynh2,dynh3,dynh4,dynhlast} and have proposed a new class of model Hamiltonians, `dynamic Hubbard models', to take it into account.
The magnitude of this effect increases as the net   charge of the ion
decreases, as discussed  e.g. in ref. \cite{dynh1}. Hence this effect will be large for $O^=$ anions \cite{tang2},
 e.g. compared to the isoelectronic $F^-$ and $Ne^0$ ions. 
As emphasized  by Bussmann-Holder et al, \cite{holder,bilz}, the oxygen $O^=$ is unstable in free space and it is stabilized in the solid by long-range Coulomb
forces only, resulting in a high polarizability \cite{over}.

In the undoped situation the oxygen ion is nominally $O^=$, which suggests that it has
a large amount of excess negative charge. The $p \sigma$ orbitals share electrons with neighboring $Cu^{++}$ ions, hence the total excess negative charge
in the $O^=$ ions is not $2$ but somewhat less, approximately $1.5$, still large. Regarding the individual $p$ orbitals,  the oxygen $p \pi$ orbitals in the plane as well as the $p_z$ orbitals
 are doubly occupied and hence will be enlarged by Coulomb repulsion. Instead, the oxygen $p \sigma$ orbitals  are less enlarged because they share their electrons
 with the neighboring $Cu$ ions.
 
As mentioned earlier, experiments show that when the  system is doped with holes, i.e. when electrons are removed from the system, these electrons come from planar orbitals.
Let us assume that an electron from a $p \pi$ orbital is removed, as shown in Fig. 4. As a consequence,  the size of the orbital is  
reduced, because it goes from being doubly occupied and hence enlarged by Coulomb repulsion to being singly occupied. This has two important consequences:
(i) the overlap with neighboring orbitals will change, changing the effective mass and  bandwidth of carriers in this band because of Franck-Condon 
overlap matrix elements \cite{tang2}, and (ii) because of the energy lowering
caused by orbital relaxation, the energy cost in removing this electron will be smaller than what would be predicted from the band structure shown in Fig. 2.
 
  \begin{figure}
\resizebox{8.5cm}{!}{\includegraphics[width=7cm]{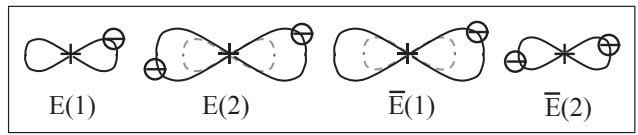}}
\caption { States with one and two electrons in an atomic p-orbital. $E(1)$ and $E(2)$ are the energies of the lowest energy states with one and two electrons respectively.
The orbital expands when the second electron is added. $\bar{E}(1)$ is the energy of one electron in the expanded orbital, and $\bar{E}(2)$ is the energy of two electrons
in the unexpanded orbital.
}
\label{figure1}
\end{figure} 
 
 As a simple illustration, consider the states of one and two electrons in an atomic p-orbital shown in Fig. 5. The wavefunction for one electron is
 \beq
 \varphi_\alpha(r,\theta)=(\frac{\alpha^5}{\pi})^{1/2}rcos \theta e^{-\alpha r} .
 \eeq
The single electron energy is 
 \beq
 E(1)=\alpha^2-\alpha Z
 \eeq
 with $Z$ the ionic charge, and the Coulomb repulsion for two electrons in this orbital is
 \beq
 U=\frac{501}{640} \alpha
 \eeq
 all in atomic units (length in units of $a_0$, energy in units $e^2/(2a_0)$).  For a single electron in the orbital the orbital exponent is
 \beq
 \alpha=\frac{Z}{2}
 \eeq
 and the atomic energy is
 \beq
 E(1)=-\frac{Z^2}{4}.
 \eeq
 For  two electrons in the orbital, the energy is minimized by the orbital exponent
 \bmath
 \beq
 \bar{\alpha}=\frac{Z}{2}-\delta
 \eeq
 \beq
 \delta=\frac{501}{2560}
 \eeq
 \emath
 reflecting the expansion of the orbital, and the energy in the two-electron atom is
 \beq
 E(2)=-2[\frac{Z}{2}-\delta]^2 .
 \eeq
 The overlap matrix element between the wavefunctions for the expanded and non-expanded orbital is
 \beq
 S=\int d^3r \varphi^*_\alpha(\vec{r} )\varphi_{\bar{\alpha}}(\vec{r})=
 \frac{(\alpha \bar{\alpha})^{(5/2)}}     {(\frac{\alpha+\bar{\alpha}}{2})^5}
 \eeq
 and becomes arbitrarily small for $Z$ approaching $2\delta=0.39$. 
  The single electron energy is higher in the expanded orbital
 \beq
 \bar{E}(1)=-[(\frac{Z}{2})^2-\delta^2]
 \eeq
 but the electrons pay this price in order to minimize the total energy. If the orbital is not allowed to expand, the total energy of the two-electron orbital is
 \beq
 \bar{E}(2)=-Z[\frac{Z}{2}-2\delta]
 \eeq
 larger than Eq. (7) by
 \beq
 \bar{E}(2)-E(2)=2\delta^2
 \eeq
 which is twice the cost in single-electron energy in expanding the orbital:
 \beq
 \bar{E}(1)-E(1)=\delta^2      .
 \eeq
 Thus, we can define the {\it relaxation energy per electron} as
 \beq
 \epsilon_R=\frac{1}{2}(\bar{E}(2)-E(2))=\bar{E}(1)-E(1)   .
 \eeq
 This  is both  the $reduction$ in energy per electron achieved by expanding the orbital versus keeping it unexpanded when a second electron is put in,
as well as the $excess$ energy of the electron remaining in the orbital after the second electron is removed if it does not relax to the unexpanded orbital.
 
 We can relate this relaxation energy to the bare and effective Coulomb repulsions. The bare Coulomb repulsion is the Coulomb energy if the orbital is not 
 allowed to expand:
 \beq
 U_{bare}=\bar{E}(2)-2E(1)=2\delta Z
 \eeq
which is the same as Eq. (3) with unexpanded orbital exponent  $\alpha=Z/2$. The ``effective $U$'' is given by
\beq
U_{eff}=E(2)-2E(1)=2\delta Z-2\delta^2      
\eeq
so that 
\beq
\epsilon_R=\frac{1}{2}(U_{bare}-U_{eff})   .
\eeq
For a single p-orbital this yields for the relaxation energy
\beq
\epsilon_R=\delta ^2=0.52eV  .
\eeq

However, this estimate does not take into account the presence of other electrons in the atom. In particular, when the $p\pi$ orbital expands the Coulomb interaction energy of an electron
in that orbital with electrons in other p-orbitals is also reduced, so that the relaxation energy should be substantially larger than Eq. (17).
We can obtain a quantitative estimate for oxygen ions using Eq. (16). The effective U for two electrons in the oxygen ion $O^n$ is given by
\beq
U_{eff}(O^n)=E(O^n)+E(O^{n+2})-2E(O^{n+1})
\eeq
with $E(O^n)$ the electronic energy for the ion $O^n$. This can be obtained from the difference in ionization energies
\beq
U_{eff}(O^n)=I(n+2)-I(n+1)
\eeq
with 
\beq
I(n)=E(O^n)-E(O^{n-1})
\eeq
the n-th ionization energy for $n\leq 1$, and $I(0)$, $I(-1)$ the first and second electron affinities of $O^0$. 

\begin{table}
\caption{ 
 n-th ionization energy $I(n)$ and effective electron-electron repulsion $U_{eff}(O^{n-1})=I(n+1)-I(n)$ 
for two electrons in $O^{n-1}$ ion.    }
\begin{tabular}{l | c | c   }
n & $I(n) (eV) $  & $U_{eff}(O^{n-1})(eV)$  \cr
 \hline
-1 & -8.75 & 10.20 \cr
0  & 1.45& 12.17   \cr
1 & 13.62 & 21.5 \cr
2 & 35.12 & 19.82 \cr
3 & 54.94 & 22.47 \cr
4 & 77.41 &  \cr
  \hline
 \end{tabular}
\end{table}

The experimental values for ionization energies and electron affinities  and resulting effective $U$'s are given in table I. Note that the effective U's for two electrons
in the $O^0$, $O^+$ and $O^{++}$ ions are all of order $20eV$or slightly larger. This is also approximately the value
for Slater's $F^0(2p,2p)$ parameter, the spherically averaged Coulomb repulsion for two electrons in p-orbitals in $O^0$
\beq
F^0(2p,2p)=\frac{93 \mu c}{128}=21.94 eV
\eeq
with $2\mu c=4.44$, from tables 15-6 and 15-7 in ref. \cite{slater}. When the ion becomes negatively charged the orbitals expand and the effective
$U$ decreases substantially, to $12.17eV$ and $10.20 eV$ for two electrons in $O^-$ and in $O^=$ respectively. 
Thus, we conclude that the `bare U' for two electrons in the neutral oxygen atom $O^0$ is of order $20 eV$ and the `effective U' for two electrons in the
negative anion $O^=$ is of order $10 eV$. According to Eq. (16) this then yields
\beq
\epsilon_R\sim 5eV
\eeq
for the lowering of energy $per$ $electron$ due to the orbital expansion when an electron is added to $O^-$ to form $O^=$. Conversely,
when adding a hole to $O^=$, i.e. removing an electron from $O^=$, the energy of the final state is $\sim 5eV$ $lower$ than would be estimated if
the orbital relaxation effect is not taken into account.

In concluding from Figure 2 that when the cuprate plane is doped with holes the oxygen $p\pi$ band remains full the effect of orbital relaxation is $not$ taken into account. 
Even a new band structure calculation with a low level of doping will not capture this effect, since it will only take into account the $average$ charge distribution
which will not change much for low doping compared to the undoped state. Instead we argue  that even doping with a single hole changes the
outcome qualitatively when the orbital relaxation effect is taken into account.

  \begin{figure}
\resizebox{8.5cm}{!}{\includegraphics[width=7cm]{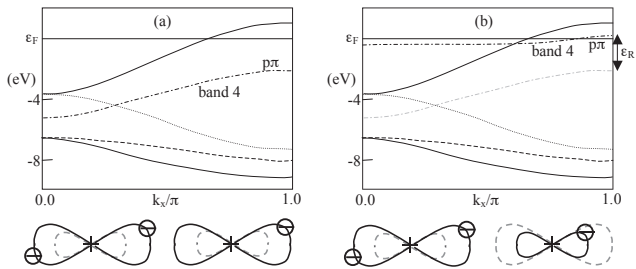}}
\caption { Effect of orbital relaxation on the band structure ((b)) (schematic) compared to the case when orbital relaxation is not taken into
account ((a)). The energies of the states of band 4, arising from overlap of oxygen $p\pi$ orbitals, are lifted  
 by an amount given by the orbital relaxation energy $\epsilon_R$ with respect to the energies when orbital relaxation is
not taking into account. In addition, the bandwidth of band 4 becomes significantly smaller due to the modulation of the hopping amplitude by the
overlap matrix elements.
}
\label{figure1}
\end{figure} 

Thus, the conclusion from the band structure shown in Fig. 2 that band 4 remains full when the system is doped with holes needs to be reconsidered. Let
$E_{initial}$ and $E_{final}$ be the initial and final energies of the system upon bringing an electron from the top of band 4 to the Fermi energy, and
\beq
\Delta=E_{final}-E_{initial}
\eeq
the energy cost of this. According to the band structure shown in Fig. 2,
\beq
\Delta\sim 2.1 eV
\eeq
and other calculations yield estimates in the range $1eV$ to $2.5 eV$\cite{bs5,mms}. These calculation (using density functional or tight binding methods) assume the 
charge distribution and kinetic, potential and interaction energies of electrons in the system {\it when the system is undoped}, and do not change upon
infinitesimal doping since on average the charge distribution doesn't change. Thus, they do not take into account the relaxation of the orbital when
an electron is removed. A correct calculation taking this effect into account would yield instead for the final energy
\beq
E'_{final}=E_{final}-\epsilon_R
\eeq
since the atomic energy is lowered by $\epsilon_R$ through the process of orbital relaxation. As a consequence, 
\beq
\Delta '=E'_{final}-E_{initial}=\Delta - \epsilon_R
\eeq
and it will change sign from positive to negative if $\epsilon_R>\Delta$, which is the case for the cuprate superconductors according to the estimates
discussed above. This then implies that when the insulating system is doped with holes, the holes will occupy the oxygen $p\pi$ band, or in other
words that the states near the top of the band of band 4 rise to the Fermi level. An additional important effect is that the bandwidth will become significantly
smaller because the hopping amplitude is reduced by the square of the overlap matrix element $S$ between  expanded and unexpanded atomic states
Eq. (8)  \cite{dynh1}.
The resulting band is shown  schematically in Figure 6.

\section{dynamic hubbard model calculation}

To take into account the effect of orbital relaxation quantitatively we use a dynamic Hubbard model \cite{dynh1,dynh4} and diagonalize it exactly for the small cluster shown in Figure 7,
with four Cu sites with one d-orbital each  and four O sites with a $p\sigma$ and a $p\pi$ orbital each.  There is in addition  a spin 1/2 degree of freedom 
associated with each oxygen $p\pi$ orbital to represent the orbital expansion/contraction. We denote by $d_{i\sigma}^\dagger$, $c_{i\sigma}^\dagger$
and $p_{i\sigma}^\dagger$ the creation operators for electrons in orbitals $d$, $p\sigma$ and $p\pi$ respectively, and
$n_{di}$, $n_{ci}$ and $n_{pi}$ the corresponding electronic site occupations. The Hamiltonian is given by
\bmath
\beq
H=\sum_i \epsilon_d n_{di}+\sum_i \epsilon_c n_{ci}+\sum_i h_i+H_{kin}
\eeq
\beqn
&&H_{kin}=t_d\sum_{ij\sigma}(d_{i\sigma}^\dagger c_{j\sigma}+h.c.)+t_1\sum_{ij\sigma}(c_{i\sigma}^\dagger c_{j\sigma}+h.c.) \nonumber \\
&+&t_1\sum_{ij\sigma}(p_{i\sigma}^\dagger p_{j\sigma}+h.c.)+t_2\sum_{ij\sigma}(c_{i\sigma}^\dagger p_{j\sigma}+h.c.)
\eeqn
\beqn
h_i&=&\epsilon_c+\epsilon_0(n_{pi\uparrow}+n_{pi\downarrow})+\omega_0\sigma_x^i+g\omega_0\sigma_z^i \nonumber \\
&+&(U_p-2g\omega_0\sigma_z^i)n_{pi\uparrow}n_{pi\downarrow}   
\eeqn
\beq
\epsilon_c=\omega_0\sqrt{1+g^2} .
\eeq
\emath
$h_i$ is the site Hamiltonian for the oxygen $p\pi$ orbital with the auxiliary spin degree of freedom to describe its expansion/contraction. The physics of this model
is discussed in detail in ref. \cite{dynhlast}.  The ground state energy of the Hamiltonian $h_i$ when the orbital is doubly occupied is
\beq
E(2)=2\epsilon_0+U_p .
\eeq

  \begin{figure}
\resizebox{8.5cm}{!}{\includegraphics[width=7cm]{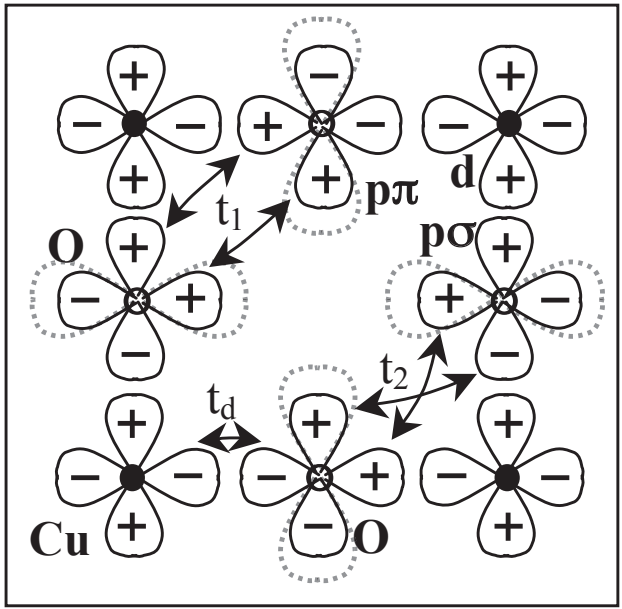}} 
\caption { 
8-site cluster for the exact diagonalization calculation. There is one  electronic  $d$-orbital at each of the four Cu sites, a $p\sigma$ and a $p\pi$ orbital at each
of the four O sites, and an auxiliary spin degree of freedom
at each O site to describe the expansion/contraction of the $p\pi$ orbitals indicated by the dotted lines.
}
\label{figure1}
\end{figure}

If we take an electron out of this orbital and don't let the orbital relax the energy is \cite{dynhlast}
\beq
\bar{E}(1)=\epsilon_c+\epsilon_0+\frac{g^2-1}{\sqrt{1+g^2}}\omega_0   .
\eeq
Therefore, we choose the parameters in the Hamiltonian $h_i$ so as to give
\bmath
\beq
E(2)=2\epsilon_{p\pi}
\eeq
\beq
\bar{E}(1)=\epsilon_{p\pi}   .
\eeq  
\emath
If the orbital is allowed to relax, the lowest energy of $h_i$ with one electron is
\beq
E(1)=\epsilon_0
\eeq
so that the relaxation energy is
\beq
\epsilon_R=\bar{E}(1)-E(1)=\omega_0\sqrt{1+g^2}+\frac{g^2-1}{\sqrt{1+g^2}}\omega_0 .
\eeq
The overlap matrix element betwee relaxed and unrelaxed orbitals is
\beq
S=\frac{1}{\sqrt{1+g^2}}
\eeq
which corresponds to the overlap matrix element between expanded and unexpanded orbitals Eq. (8), 
and determines the bandwidth of the renormalized band when the orbital relaxation effect is taken into account. We take it here as a free parameter. For given
$S$ and relaxation energy $\epsilon_R$ the parameters in the Hamiltonian $h_i$ are then
\bmath
\beq
g=\sqrt{\frac{1}{S^2}-1}
\eeq
\beq
\omega_0=\frac{\epsilon_R}{\sqrt{1+g^2}+\frac{g^2-1}{\sqrt{1+g^2}}}
\eeq
\beq
\epsilon_0=\epsilon_{p\pi}-\epsilon_R
\eeq
\beq
U_p=2\epsilon_R    .
\eeq
\emath
Finally, to match the band structure results of Figure 2 we take $\epsilon_d=-5.2$, $\epsilon_{p\sigma}=-5.5$, $\epsilon_{p\pi}=-4.7$, 
$t_1=0.65$, $t_2=0.35$, $t_d=1.75$. 

  \begin{figure}
\resizebox{8.5cm}{!}{\includegraphics[width=7cm]{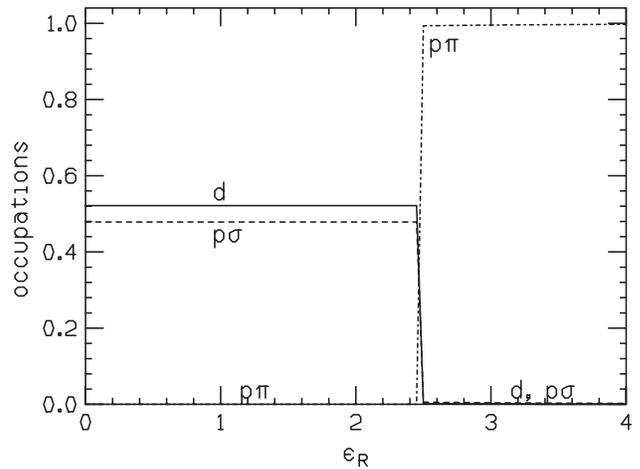}} 
\caption { 
Occupation of the orbitals in the cluster of Figure 7  in the ground state when there is one hole in the cluster as function of the orbital relaxation energy $\epsilon_R$  (in eV) 
for $S=0.333$. 
}
\label{figure1}
\end{figure}

  \begin{figure}
\resizebox{8.5cm}{!}{\includegraphics[width=7cm]{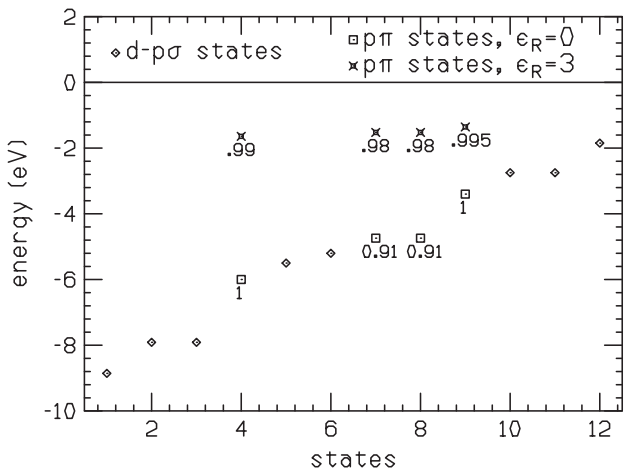}} 
\caption { 
Energy eigenvalues for the cluster of Fig. 7 with one hole in the absence of orbital relaxation and  when orbital relaxation is  included with  the dynamic Hubbard model, with
$S=.333$ and  $\epsilon_R=3$. The four $p\pi$ states in the cluster are moved up in energy by several $eV$
when orbital relaxation is included. The numbers next to the symbols indicate
the $p\pi$ occupation for the state. Note also that the spread in energy of the $p\pi$ states becomes much smaller in the
presence of orbital relaxation.
}
\label{figure1}
\end{figure}

The Hilbert space for one hole in this cluster has 192 states (12 states for the hole and 16 states for the auxiliary spins) and thus the Hamiltonian
Eq. 27 is easily diagonalized in this subspace. We calculate the occupation of the orbitals $d$, $p\sigma$ and $p\pi$ as function of the
relaxation energy $\epsilon_R$ and plot the occupations for the lowest energy state for the case $S=0.333$ in Figure 8. 

The results are qualitatively as expected. For zero or small relaxation energy the hole resides in the $d$ and $p\sigma$ orbitals, i.e band 5 in
Figure 2. Note in Fig. 8  that the occupation of the 
$d$ orbitals is slightly larger than that of the $p\sigma$ orbitals, contrary to the results shown in Figure 3, lower right panel. This is simply due to
the finite size of the cluster used. When the relaxation energy exceeds a critical value $\epsilon_{Rc}\sim 2.5 eV$ the hole occupies
the oxygen $p\pi$ orbital, i.e. the top of band 4 in Figure 2. In other words, the states at the top of that band are now at the Fermi energy, as shown
schematically in Figure 6, right panel.

The critical value $\epsilon_{Rc}\sim 2.5 eV$ is similar to the distance between the top of band 4 and the Fermi energy in Fig. 2, or between the top of band 4 and
the top of band 5 in Fig. 2. This critical value depends somewhat on the value assumed for the overlap matrix element $S$, and decreases monotonically from
$\epsilon_{Rc} = 2.7 eV$ to $\epsilon_{Rc} = 1.6 eV$ for $S$ going from $0$ to $1$.

In Fig. 9 we show the energy eigenvalues for all the electronic states with one hole  in the cluster in the absence of orbital relaxation,
and the change in the energy of the $p\pi$ states when the effect of orbital relaxation is included. The  diamonds and squares denote the
states in the absence of orbital relaxation, of $d-p\sigma$ and $p\pi$ character respectively, as determined by the occupations of the sites.
 For the states denoted
by the diamonds the $p\pi$ occupation is zero or very small. Note that the highest energy states in the absence of orbital
relaxation are diamonds, corresponding to the $d-p\sigma$  states of band 5 in Fig. 2, which are the states where holes would be
created in the absence of orbital relaxation. When orbital relaxation is included, the $p\pi$ states (squares) go up in
energy and become the states denoted by the crosses, which are higher in energy than the highest $d-p\sigma$ states,
indicating that doped holes will be created in those states. The numbers next to the squares and crosses give the hole occupation of the $p\pi$ orbitals for those states.
 Note also that the spread in energy of the states 
denoted by the crosses is much smaller than the states denoted by the squares, reflecting the band narrowing caused
by orbital relaxation. Examination of the states denoted by the crosses shows that they are small `electronic polarons',
with the  auxiliary spin distortion representing the expansion/contraction of
the orbital following the occupation of the $p\pi$ orbitals as expected.

\section{consequences for normal state properties and for superconductivity}

If the doped holes in the cuprate superconductors occupy the O $p\pi$ rather than the O $p\sigma$ orbitals this has
fundamental  consequences for the understanding of both the normal state of the cuprates and the
mechanism of superconductivity. In particular, Goddard et al \cite{goddard2}, Stechel and Jennison \cite{sj} and Birgenau et al \cite{birg} 
 have proposed
superconductivity mechanisms based on magnetic interactions between oxygen $p\pi$ hole carriers and Cu spins, and Ikeda \cite{ikeda} has proposed
an exciton mechanism for pairing of
O $p\pi$ holes induced by interactions with $d-p\sigma$ excitations. 

Here   we focus  on the physics predicted by the theory of hole superconductivity \cite{hole}.  The hole carriers doped into the O $p\pi$ states shown in Fig. 9 that
were raised to the Fermi energy by orbital relaxation will be highly dressed by the orbital relaxation processes, with quasiparticle weight given by \cite{undr,undr2}
\beq
z=S^2
\eeq
for infinitesimal hole doping.  If we assume that half of the missing electron in the $Cu-Op\sigma$ bond resides on the $Cu$ and half on the $O$ it corresponds to $Z=0.5$
in Eq. (2), yielding
 $\bar{\alpha}=0.054$ in Eq. (6)   and $S=0.263$ from Eq. (8). If we assume the $d^{9.45}$ occupation for the $Cu$ ion predicted by band structure calculations \cite{mms}
 it corresponds to $Z=0.45$, $\bar{\alpha}=0.029$ and $S=0.106$. These estimates illustrate that the value of $S$  (and hence $z$)  is very sensitive to the charge distribution but
 in any event is likely to be much smaller than unity.

As the doping level increases, the quasiparticle weight increases as \cite{undr2}
\beq
z(n)=S^2[1+\frac{n_h}{2}\Upsilon]^2
\eeq
with $n_h$ the hole concentration and  $\Upsilon$ the `undressing parameter'\cite{undr2}
\beq
\Upsilon=\frac{1}{S}-1
\eeq
which should be particularly large for negative ions such as $O^=$. This will give rise to a vanishingly small quasiparticle weight for infinitesimal hole doping
($z=0.106^2=0.01$ if we use the $d^{9.45}$ value)  that
grows upon doping leading to increasing coherence in the normal state. This behavior is seen experimentally \cite{hong,shen} but attributed
to physics of the $t-J$ model and the Mott-Hubbard gap in the O $d-p\sigma$ band instead of the physics and band discussed here.  
In addition, within our theory the same parameter $\Upsilon$ determines the correlated hopping interaction $\Delta t$  that gives rise to pairing and
superconductivity \cite{hole2} driven by lowering of kinetic energy \cite{kinenergylowering}
\beq
\Delta t=\Upsilon t_h
\eeq
with $t_h=t S^2$ and $t=t_{\pi \pi}=t_1$ the bare $p\pi-p\pi$ hopping. Through the same physics, the model predicts increased coherence  \cite{undr,undr2} and increased
low energy optical spectral weight \cite{apparent,kinen} as the system goes superconducting, which is  seen experimentally in
photoemission \cite{hong} and optical \cite{basov,marel} experiments.

The mechanism of hole superconductivity \cite{hole} only operates when a band is almost full, hence it can only  be
relevant to the cuprates  if the doped holes go into a $p\pi$ band rather than a $p\sigma$ band that is half-full in the undoped case.
The predicted doping dependence of $T_c$ \cite{prb90}  closely resembles the doping dependence seen in the cuprates \cite{torrance}. Many other
predictions of the model are in agreement with observations \cite{prb90,kinen,tunn}. In addition the model predicts a strong tendency to charge inhomogeneity \cite{dynh3} as seen 
experimentally.

\section{discussion} 
The question of the nature of the charge carriers in the high $T_c$ cuprates is undoubtedly a complicated one, both because of the existence of strong correlations induced by the
Hubbard $U$ on the Cu sites \cite{anderson} and because of the effects of strong orbital relaxation in the negative $O^=$ anions focused on in this paper. Band structure calculations
capture neither of these effects. For the past 25 years the physics community has focused its attention on the former of these effects and ignored the latter.  With this paper we attempt to restore some balance to this situation. 

It is possible that in fact doped carriers occupy $both$ Cu-$Op\sigma$ bands as the Mott-Hubbard gap closes $and$ an $O p\pi$ band lifted to the Fermi energy
by orbital relaxation. We have considered such a two-band model \cite{twob} and found that high temperature superconductivity would also result, driven by
the kinetic pairing interaction of the $O p \pi$ carriers that would induce a weaker pairing of the $Cu-Op\sigma$ carriers.

More generally, the purpose of this paper is to point out  that the effect of orbital relaxation on the band structure {\it cannot be ignored} when negative ions are involved, and that  it is
not taken into account in standard band structure calculations. Similar physics is emphasized in the approach developed by
Fulde and collaborators \cite{fulde1,fulde2,fulde3} as  a substitute for standard band structure approaches. Within dynamical mean field theory \cite{dmft} it is also
possible to take the effects focused on in this paper into account\cite{bach}, although this has not yet been done in the context of a realistic band structure calculation.

In summary, we argue that the arguments and calculations in this paper indicate that orbital relaxation of the oxygen 
$p\pi$ orbitals in the cuprate superconductors   raises an $Op\pi$ bands to the Fermi level when holes are doped into
the system, hence that the carriers responsible for superconductivity in the cuprates are holes in a full $p\pi$ band rather than in a
half-filled  $d-p\sigma$ band as generally assumed. In a band close to full and in the presence of strong orbital relaxation the
theory of hole superconductivity predicts that high temperature superconductivity results, with many
characteristic features seen in the high $T_c$ cuprates.

To conclude we point out that a strong argument in favor of our point of view  is that if it is correct for the cuprates it is likely to also explain the high temperature superconductivity
of several other materials such as $MgB_2$ \cite{mgb2ours}, iron pnictides and dichalcogenides \cite{pnicours}, and explain the reason for why all high temperature superconducting materials
appear to have $negative$ $ions$ \cite{first,matmech}, an observation also made by A.W. Overhauser\cite{over2}. It also suggests an explanation for  why elements under high pressure are superconducting \cite{hamlin}, for the fact that electron-doped cuprates have
hole carriers  in the regime where they become superconducting \cite{dagan}, and for the pervasive presence of hole carriers in superconducting materials ranging from the
elements \cite{chapnik} to A15 compounds \cite{a15} to $MgB_2$ \cite{mgb2} to high $T_c$ cuprates.

\end{document}